\documentclass[
nofootinbib,
 amsmath,amssymb,
 aps,twocolumn,nofootinbib
]{revtex4-1}
\usepackage[parfill]{parskip}
\usepackage{graphicx}
\usepackage{amssymb}
\usepackage{mathtools}
\usepackage{color}

\newcommand{\be}{\begin{equation}}
\newcommand{\ee}{\end{equation}}
\newcommand{\ba}{\begin{eqnarray}}
\newcommand{\ea}{\end{eqnarray}}

\begin{document}
\title{Maximum latent heat of neutron star matter
independently of General Relativity
}
\author{Eva Lope-Oter and Felipe J. Llanes--Estrada}
\affiliation{Univ. Complutense de Madrid, dept. F\'isica Te\'orica and IPARCOS, Plaza de las Ciencias 1, 28040 Madrid, Spain.}

\begin{abstract}

We establish bounds on the maximum possible specific latent heat of cold neutron-star matter derived from hadron physics alone. 
Existing chiral perturbation theory computations for the equation of state, together with perturbative Quantum Chromodynamics, relevant at highest densities (even if they would turn out not to be physically realizable) bind the maximum latent heat which is possible in actual neutron stars. 
Because these are already near gravitational collapse in General Relativity, no denser form of cold matter can exist: thus, the bounds are a generic physical limit. 
Even in scenarios that modify the theory of gravity, the existence of a family of latent-heat maxima is relevant to diagnose progress in the knowledge of the equation of state of neutron matter, by quantifying the maximum possible (presumed) phase transition that 
its error bands would allow. Thus, latent heat is a natural benchmark for the equation of state in cold QCD.

\end{abstract}

\maketitle

\section{Latent heat\\ of first order phase transitions}

We presently ignore whether neutron-star matter undergoes a first-order phase transition to an exotic, perhaps nonhadronic phase, of great interest to nuclear and particle physics~\cite{Chesler:2019osn}. Many have been proposed, such as color-superconducting phases~\cite{Alford:2007xm}, inhomogeneous (crystalline-like) phases~\cite{Fulde:1964zz} or flavored~\cite{Oertel:2016xsn} or mixed phases~\cite{Heiselberg:1994fy} among many examples; but whether any phase is physically realized in neutron stars remains an object of both theoretical and observational investigation~\cite{Llanes-Estrada:2019wmz}.

Possible (one or more) first-order phase transitions would leave distinct observable traces, such as a kink in the mass-radius diagram (accessible when neutron star radii become more routinely measured). They would be characterized by a jump in energy density $\varepsilon_{E}-\varepsilon_{H}$ between the hadronic ($H$) and exotic ($E$) phases,  as a discontinuity in the free energy presents itself upon changing a thermodynamic variable such as temperature or density~\cite{Nayak:2012}. 
 A way to quantify the leap is the specific latent heat of the transition $L$,  normalized to the unit mass. With mechanical dimensions $E/M$, it is a pure number in natural units with $c=1$; table~\ref{tab:L} shows a few cases, from condensed matter~\cite{Kosugi:2021} to nuclear physics~\cite{Carbone:2010ut}.

\begin{table}[b]
\caption{Some salient values of latent heat in natural units. \label{tab:L} }
\begin{tabular}{|c|c|} \hline
Substance/transition  & $L$ \\ \hline 
He-3 superfluid                   & 1.5 $\mu$J/mol = $5.5\times 10^{-24}$\\ 
NdCu$_3$Fe$_4$O$_{12}$ perovskite & 25.5 kJ/kg = $2.8\times 10^{-13}$\\ 
Ice-water             & 79.7 cal/g = $3.71\times 10^{-12}$ \\
Nuclear evaporation   & 30 MeV/A = $3\times 10^{-2}$ \\
Neutron star matter? & $>O(0.1)$? \\
\hline
\end{tabular}
\end{table}

In this letter we  discuss the room for a possible first-order phase transition at zero temperature $T$ in neutron-star matter. 
The Gibbs thermodynamic-equilibrium condition determines at what critical chemical potential $\mu_c$  will the pressure of the two phases  be equal
\begin{equation}\label{Gibbscondition}
      T_H = T_E = 0 \ , \ \ \ \ 
      \mu  = \mu_c \ , \ \ \ \ 
      P_H = P_E := P_c\ .
\end{equation}
In the low-density regime the starting point is
\begin{equation}\label{def:Ln}
L|_n:=\frac{\Delta E}{NM_N}\ ,
\end{equation}
a latent heat per nucleon normalized to the vacuum neutron mass of $940$ MeV, computed from $\Delta E :=  E_E- E_H$, and to which we have added a subindex $n$ to distinguish it from $L|_\varepsilon$ defined shortly in Eq.~(\ref{LperA}).

To obtain $L|_n$ from the Equation of State (EoS) $P(\varepsilon)$ we integrate  the first law of thermodynamics in terms of the  pressure, the number density $n$ (baryon number is conserved) and the energy per nucleon ($E = \varepsilon/n$),
\begin{eqnarray}
      \label{integralLey1}
\int \frac{1}{P(E)}dE &=& \int \frac{dn}{n^2} \ ,
\end{eqnarray}
with  limits extending from the transition point $n_{\rm tr}=n_{\rm H}$ (where the phase $H$ is pure) to that $n_{E}$ where the medium is completely in the presumed exotic phase.

Other research~\cite{Lindblom:1998dp,Seidov:1971sv}, instead of $L$ (be it $L|_n$ or $L|_\varepsilon$), often discusses  the difference of the energy densities between the two phases, $\Delta \varepsilon$. While energy density is natural in treating the relativistic stress-energy tensor, as 
$T^{00}=\varepsilon$, it difficults comparing phenomena across physics domains, so we prefer to
adopt the wider convention and refer to (specific) $L$ instead of $\Delta \varepsilon$.
To relate them, note that the difference of energies per nucleon
$E_E-E_H$ can be calculated  from Eq.~(\ref{integralLey1}), that can be immediately evaluated because $P$ is constant over the phase transition. Then,
\begin{equation}\label{diff}
     \Delta E =P_H \frac{(n_E-n_H)}{n_E n_H}
\end{equation}
allows to compute Eq.~(\ref{def:Ln}) for $L|_n$ that will be shown below in figure
\ref{fig:L(n)}.

Closer in spirit to existing work in the field is the natural relativistic modification of Eq.~(\ref{def:Ln}) 
to include the nuclear (anti)binding energy,  $B$. Through  
\begin{equation}\label{Eandn}
\varepsilon= n(M_N c^2+ B/A)
\end{equation}
and Eq.~(\ref{diff}), a relativistic definition for the latent heat follows, that can be directly evaluated from the Equation of State (section~\ref{sec:eos}) $P(\varepsilon)$ as read off in  $T^{\mu\nu}$ ,
\begin{equation}\label{LperA}
     L|_\varepsilon = P_H \frac{(\varepsilon_E-\varepsilon_H)}{\varepsilon_E  \varepsilon_H} \ .
\end{equation}
At low $\varepsilon$, 
Eq.~(\ref{LperA}) is equivalent to Eq.~(\ref{def:Ln}) save for the binding energy $0\simeq (B/A) << M_N$, introducing a few percent error quantified below. 
Near the density allowing maximum $L$, Eq.~(\ref{LperA}) is a very practical definition of $L$.
Nevertheless we will offer a comparative with the quantity most directly related to the nonrelativistic limit, the $L|_n$ from Eq.~(\ref{def:Ln}) with Eq.~(\ref{diff}) substituted therein.

\section{Equations of state for neutron stars valid for both General Relativity and 
modified gravity}\label{sec:eos}

General Relativity (GR) is widely accepted as the correct theory of gravity inside neutron stars. But this requires further testing: Einstein's equations $G^{\mu\nu} = \kappa T^{\mu\nu}$ have not been exhaustively constrained at such high $\varepsilon$. For example, solar-system, binary pulsar, and gravitational-wave propagation tests are basically  {\it in vacuo}~\cite{Llanes-Estrada:2019wmz,Abbott:2020jks}.
Several constraints on  $P(\varepsilon)$
coming, for example, from the maximum mass of a neutron star or its tidal deformability, are obtained within GR.
 However, testing the theory of gravity in the presence of matter requires prior knowledge of the EoS. For this purpose, earlier work was dedicated to the nEoS sets~\cite{Oter:2019kig}~\footnote{
At {\tt http://teorica.fis.ucm.es/nEoS} the reader can download several thousand equations of state.}. They systematically map out the uncertainty in the EoS coming from hadron physics: perturbative Quantum Chromodynamics (pQCD) and the various existing chiral perturbation theory computations at low density (ChPT) as well as  first principles (thermodynamic stability and causality) alone.
The nEoS sets are less constrained than others in the recent literature~\cite{Godzieba:2020tjn}, but more reliable for testing  gravity. 

\begin{figure}[t]\vspace{-0.3cm}
\includegraphics[width=0.95\columnwidth]{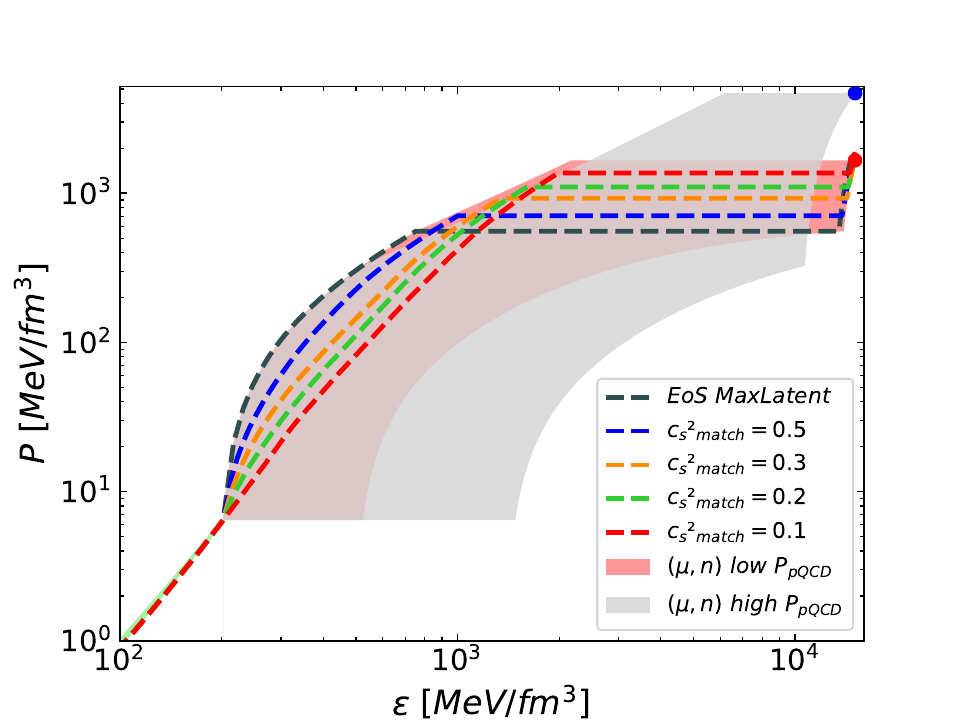}
\includegraphics[width=0.95\columnwidth]{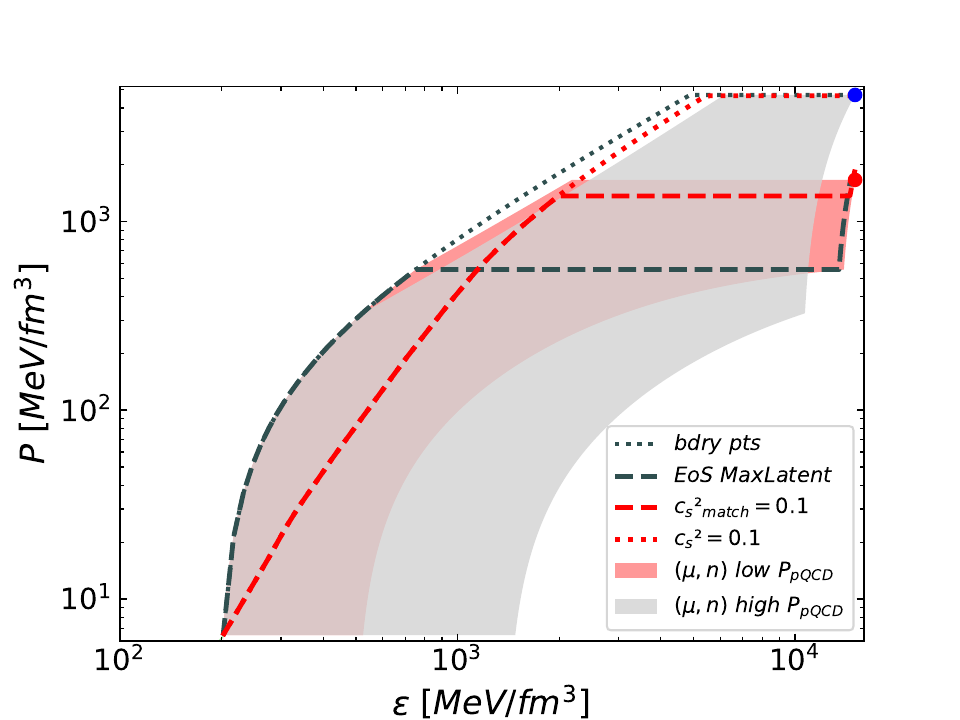}
\caption{\label{fig:band} nEoS~\cite{Oter:2019kig} theoretically allowed eq. of state of neutron matter from hadron physics alone. Between low-$\varepsilon$ ChPT and high-$\varepsilon$ pQCD (only the matching points are shown at the upper right corner, with $\varepsilon=1.5$GeV/fm$^3$ and $P\in(1.65,4.7)$ GeV/fm$^3$)~\cite{Drischler:2016djf,Kurkela:2009gj,Komoltsev:2021jzg} we extend uncertainty bands. The gray and tile-red areas correspond to the constraints from imposing simultaneously the limit for  $n$ and $\mu$  at the high-$\varepsilon$ pQCD and the low-$\varepsilon$ pQCD  point, respectively. 
{\bf Top}: Example EoS inside the band are ordered upwards by increasing $c_s^2$ at the ChPT matching point. 
 The low-density ChPT approximation is used up to $1.3n_s$, just above the nuclear saturation density.
{\bf Bottom}: The extreme case with maximum latent heat (leftmost dashed grey line) and a softer EoS with matching $c_s^2=0.1$ that will be used for later comparison. The dotted lines correspond to the same EoS without constraints from the $(n,\mu)$ plane).
 \label{fig:EoS}
 }
\end{figure}

Figure~\ref{fig:EoS} shows a small sample.
These are EoS of zero charge, $\beta$-stable neutron-star matter (NSM). 
At lowest number densities $n\leq 0.05 n_{\rm sat}$ nuclear data directly  constrains the crustal EoS~\cite{Negele:1971vb,Baym:1971pw}. (The saturation density is $n_{\rm sat}\simeq 0.16/$fm$^3$ that, depending on the order of perturbation theory taken to match at lower densities, corresponds to $B/A\simeq 16\pm 1$ MeV added to the nucleon mass, and to an energy density around 153 MeV/fm$^3$).

Low $0.05 \leq n \leq 2 n_{\rm sat}$ densities are reasonably taken care of by ChPT predictions. We have exhaustively explored the systematics due to the various groups trying different many body techniques, momentum-cutoffs for Feynman diagrams and orders in perturbation theory.  Shown here are the $N^3LO$ bands of~\cite{Drischler:2016djf,Drischler:2020yad} up to $n_m=1.3 n_{\rm sat}$ (checks have been conducted employing $n_m=2.0 n_{\rm sat}$ to explore the systematics, and barely any difference on $L|_n$ was found, although the energy density $\varepsilon$, at which the maximum transition occurs, shifts to higher values).
The highest densities, at baryon chemical potential $\mu\geq 2.6\  GeV$ and above (this corresponds to an energy density of about 15 GeV/fm$^3$), can be studied with pQCD~\cite{Kurkela:2009gj}. Such high densities are probably not reachable in neutron stars within GR, but nonetheless provide a powerful constraint to whatever shape the actual physical hadron EoS takes. 

The big unknown is the EoS for intermediate $\varepsilon$, 
extending to a maximum $P=4680$  MeV/fm$^3$ or about $\varepsilon_E=15$ GeV/fm$^3$ (where pQCD is naturally assumed to hold).
Unlike in earlier works, this region depends on the precise QCD matching-point pressure, whether at the top or at the bottom of the allowed interval~\cite{Lope-Oter:2021uft} (we depict these two extremes with different tonalities in the figure). They have been obtained by enforcing that the derivative of the EoS curves satisfy $0\leq P'(\varepsilon):= c^2_s \leq 1=c^2$ (that is, respecting monotony and causality). 
Changes respect to earlier works are due to new integral constraints from causality in the $(n,\mu)$ plane have been brought to bear~\cite{Komoltsev:2021jzg}. These make a planar representation difficult, since those constraints depend on the chemical potential at the pQCD matching point and are laborious to bring to the $(\varepsilon,P)$ plane shown here and natural for the stress-momentum tensor $T^{\mu\nu}$.
But broadly speaking, and refering to the bottom plot of Fig.~\ref{fig:band}, either the gray band (for EoS entering pQCD at the highest point in the upper right corner) or the tile-red band (for those entering at the lowest point, red online) can contain the (unknown) physical EoS.

We sample the contained region with a  500$^2$ grid.  Any $P(\varepsilon)$ in this region is matched to ChPT at a number density $n_m$, with  slope $P'=c^2_{s,m} \in (0,1)$ (given in the figure legends). At each successive grid point, a   random-slope step within those limits is taken.

Additionally to the different ChPT matching point, the few curves selected for the top and bottom plots of figure~\ref{fig:EoS}  can be qualitatively distinguished.
Those in the top plot have smoothly increasing slope $P'$ up to $c^2=1$, then flatten at $P'=0$ as if a first order phase transition took place. Typically, maximum latent heats will be found among this class.

The bottom plot is used to show the transition that we find to have the largest possible latent heat (dashed line to the left, following the causality limit $c=1$ to the maximum point allowed by  the $n$, $\mu$ constraints for low-$\varepsilon$ pQCD ~\cite{Komoltsev:2021jzg}  or red area). The softer EoS to its right (red online) has a less extreme latent heat value that ressembles those in other works~\cite{Jokela:2020piw}.

\section{Bounds on the latent heat of neutron stars} 

\subsection{Numerical computation of $L|_\varepsilon$}

All the tools are now ready to discuss numerical values for the specific latent heat $L$, starting by the simpler $L|_\varepsilon$. We proceed by following each EoS that is compatible with all theoretical requirements (monotony and causality at every point, and satisfaction of both ChPT and pQCD constraints in their domains of validity), one at a time, from lower to higher $\varepsilon$.

At each grid point we ask ourselves what is the maximum stretch of $P'=0$ (first order phase transition) that could take place without violating any of those requirements, that is, we momentarily assume that very grid point to be the lower end of a phase transition, $(\varepsilon_H, P_H)$.

\begin{figure}
\includegraphics[width=0.9\columnwidth]{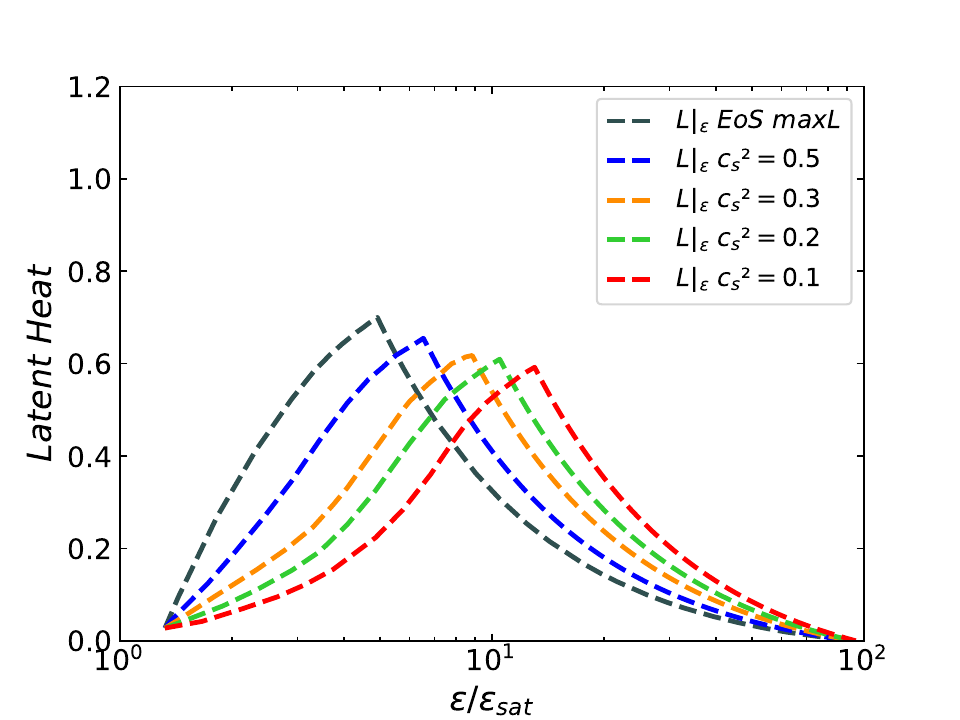}
\caption{Maximum specific latent heat $L|_\varepsilon$  with eq. of state that fits within the grey nEoS band from figure~\ref{fig:EoS}. The $OX$ axis (energy density at which the phase transition triggers) extends to 100$n_{\rm sat}$  (at chemical potential $\mu_B=2.6$ GeV) where  pQCD is matched. 
An absolute maximum $L|_\varepsilon \simeq 0.7$ is reached for
$\varepsilon \simeq 5 \varepsilon_{\rm sat}$.
. \label{fig:L}}
\end{figure}

The latent heat for such phase transitions, extending in $\varepsilon_E$ as far as possible towards high-density pQCD or the causality band limit in figure~\ref{fig:EoS} (note that its apparent occasional  steepness is due to the different $OX$ and $OY$ axes log scales), are calculated and taken to figure~\ref{fig:L}.
The largest maximum latent heat that we find  possible with the current QCD understanding, 
$L\simeq 0.7$, would be reached for $\varepsilon \simeq 5 \varepsilon_{\rm sat}$, (for an EoS matched to nuclear matter  with maximum slope $c^2_s\simeq 1$ at 1.3 $n_{\rm sat}$).

But if the phase transition is specified to trigger at an energy density $\varepsilon_H$ (number density $n_H$) by whatever physical mechanism, the maximum  possible $L|_\varepsilon$ will be the corresponding point along the top curve (and thus, smaller than the maximum). If, moreover, $P$ at the transition point is also specified, such as {\it e.g.} because it must lie on a specific low-density EoS, then the maximum $L|_\varepsilon$ that QCD theory allows will lie on one of the lower curves. 

\begin{figure}[h]
\includegraphics[width=0.9\columnwidth]{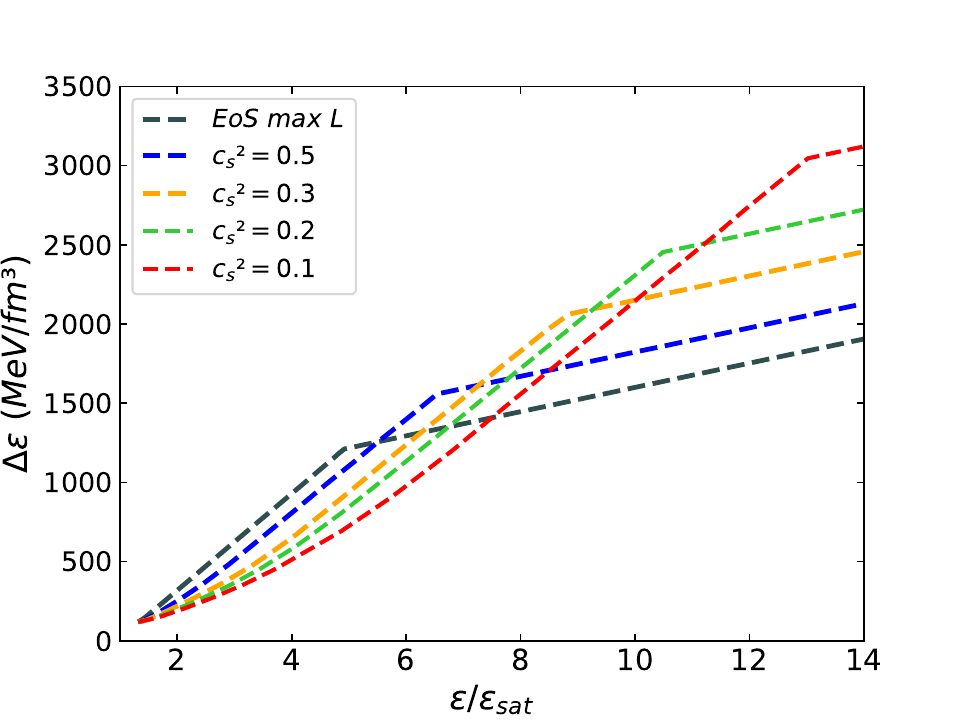}
\caption{\label{fig:maxDE}  Maximum  $\varepsilon$ discontinuity of Eq.~(\ref{Seidovsjump}) allowed in GR for phase transitions with the EoS  of from figure~\ref{fig:EoS} (top panel) and~\ref{fig:L}.}
\end{figure}

Because of simplicity, the literature often discusses a maximum ``critical'' discontinuity  in the energy density, the length of a zero-slope stretch in the $(\varepsilon,P)$ diagram. Seidov's small-core approximation~\cite{Seidov:1971sv}  would yield
\begin{equation}\label{Seidovsjump}
\Delta \varepsilon := \varepsilon_E-\varepsilon_H  = \varepsilon_H \left( \frac{1}{2} +\frac{3}{2}\frac{P_H}{\varepsilon_H} \right)
\end{equation}
whose maximum possible value, for comparison with that body of work, is shown in figure~\ref{fig:maxDE}. The EoS correspond to those in figure~\ref{fig:EoS} and are the same as used to produce figure~\ref{fig:L}, focusing in the region of interest for neutron stars within General Relativity. 

Having shown what is known with certainty from hadron physics alone, we can now turn to a comparison with the traditional Seidov bound, that assumes that GR holds 
(other specific theories of gravity would provide analogous results).

This follows from Seidov's observation that material added to the star, while its core undergoes a phase transition, accrues to the mass without increasing $P$, moving closer to black-hole gravitational collapse.
Since the EoS in the hadronic phase relates $P_H$ and $\varepsilon_H$, we can think of Eq.~(\ref{LperA}) as providing a function 
$L_{\rm Seidov}=L_{\rm Seidov}(\Delta \varepsilon,\varepsilon_H)$.

\begin{figure}[h]
\includegraphics[width=0.9\columnwidth]{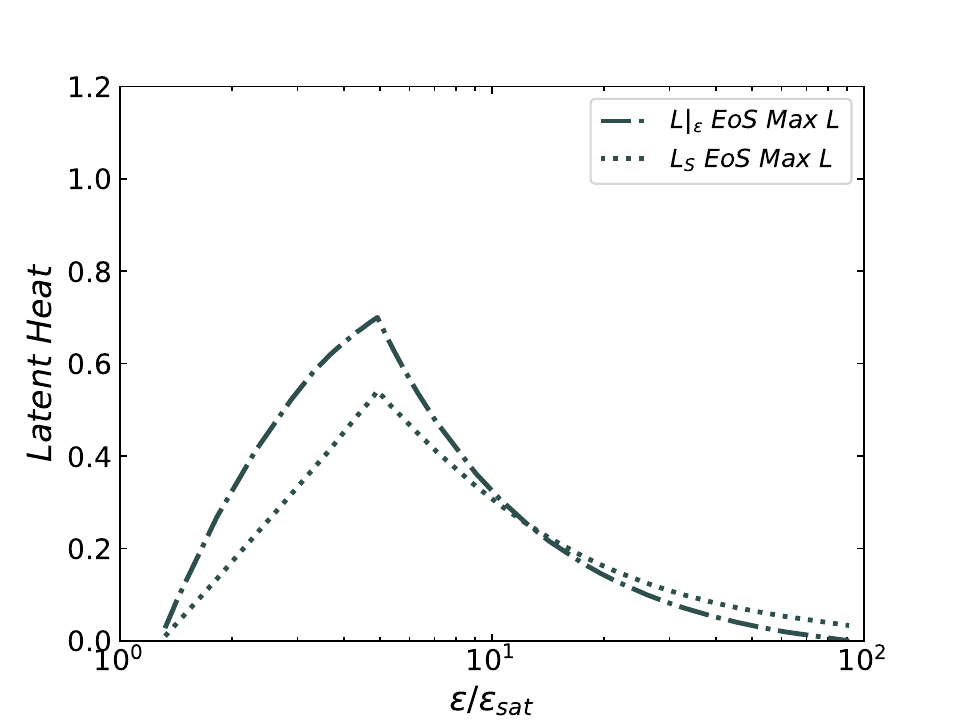}
\caption{We compare the bound on the latent heat $L|_\varepsilon$ (upper line, dashed) from microscopic hadron physics alone (see fig.~\ref{fig:L}) with the maximum latent heat that a static body in equilibrium can tolerate in General Relativity, the Seidov limit~\cite{Seidov:1971sv}  (bottom line, dotted). 
\label{fig:compareSeidov}}
\end{figure}

We plot $L_{\rm Seidov}$,  together with the top $L|_\varepsilon$ line of figure~\ref{fig:L}, in figure~\ref{fig:compareSeidov}.
For large swaths of density (including those relevant for neutron stars in GR), the purely hadronic calculation is less tight than the Seidov limit. This, however, cannot be used with modified gravity theories, so when handling them our less constraining limit becomes relevant. A good diagnostic for future progress in the EoS would be to drastically cut the separation between the two curves. A promising new approach employs the functional renormalization group to bring down the point of matching to QCD~\cite{Leonhardt:2019fua} from the high densities needed for perturbation theory.

It should also be noted that the lower line of Seidov's limit entails a perturbative assumption about the exotic core of the star being of small size, so a full numerical computation might displace it. Our hadron limit, the upper curve, is (numerically) exact.

\subsection{Numerical computation of $L|_n$}

\begin{figure}[h]
\centering
\includegraphics[width=\columnwidth]{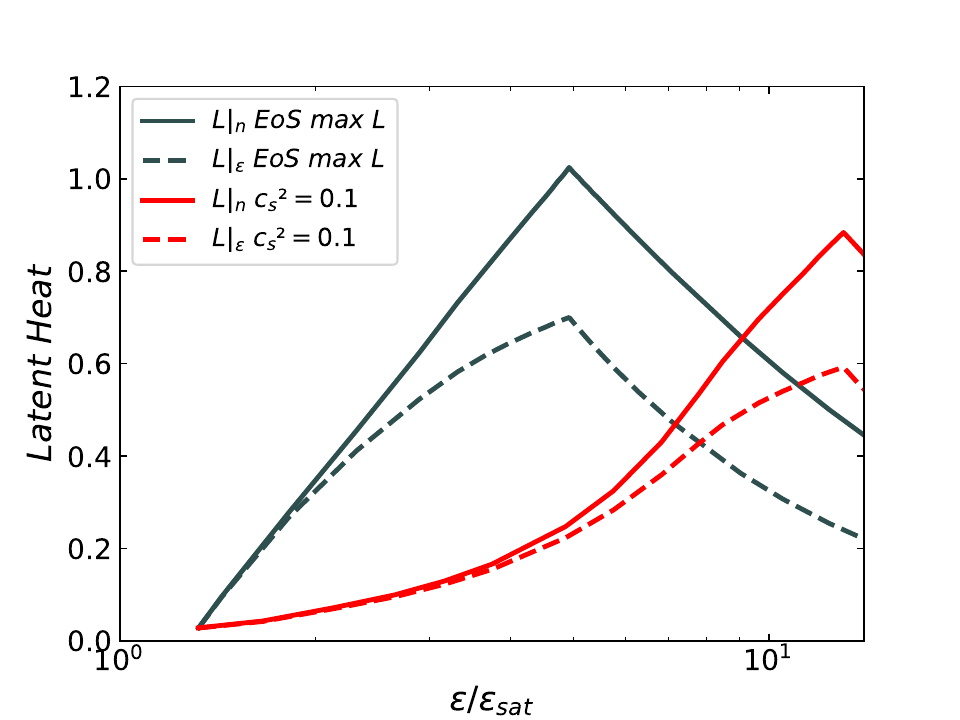}
\caption{\label{fig:Lfinal}
Solid lines: maximum possible latent heat $L|_n$ (top)  
and for a soft $c_s^2=0.1$ eq. of state in the allowed nEoS band, as function of the starting point $\varepsilon_H$ where the transition triggers. Dashed lines: same for $L|_\varepsilon$. The top two lines are absolute bounds on the respective latent heats allowed by the strong interactions.
}
\end{figure}

The difference between computing $L|_n$ and  $L|_\varepsilon$ (shown in fig.~\ref{fig:Lfinal}) grows with energy density. Around $\varepsilon\simeq 2.6 \varepsilon_{\rm sat}$, the stiffest EoS leads to a 15\% change relative to  $L|_\varepsilon$; with the softest EoS at the bottom of the allowed band, the two heats agree within 15\% all the way to $\varepsilon\simeq 5.8 \varepsilon_{\rm sat}=5.8 \times 153$MeV/fm$^3$.

The corresponding binding energy per nucleon $B/A$ up to which the relative difference between $L|_n$ and  $L|_\varepsilon$ 
remains within 15\% is about $ 20\% M_N$ (with the precise figure depending on the EoS), still 
in a regime where the total energy is mass-dominated.

Beyond such energy densities it becomes really necessary to distinguish
$L|_\varepsilon$ from $L|_n$. 
The first one is quite straightforwardly obtained from Eq.~(\ref{LperA}).  
To compute the second it is necessary to solve
for the relation between $n$ and $E$ in Eq.~(\ref{Eandn}), that requires the binding energy per nucleon at each point of the grid, $(B/A)_i$.

\begin{figure}
\includegraphics[width=0.9\columnwidth]{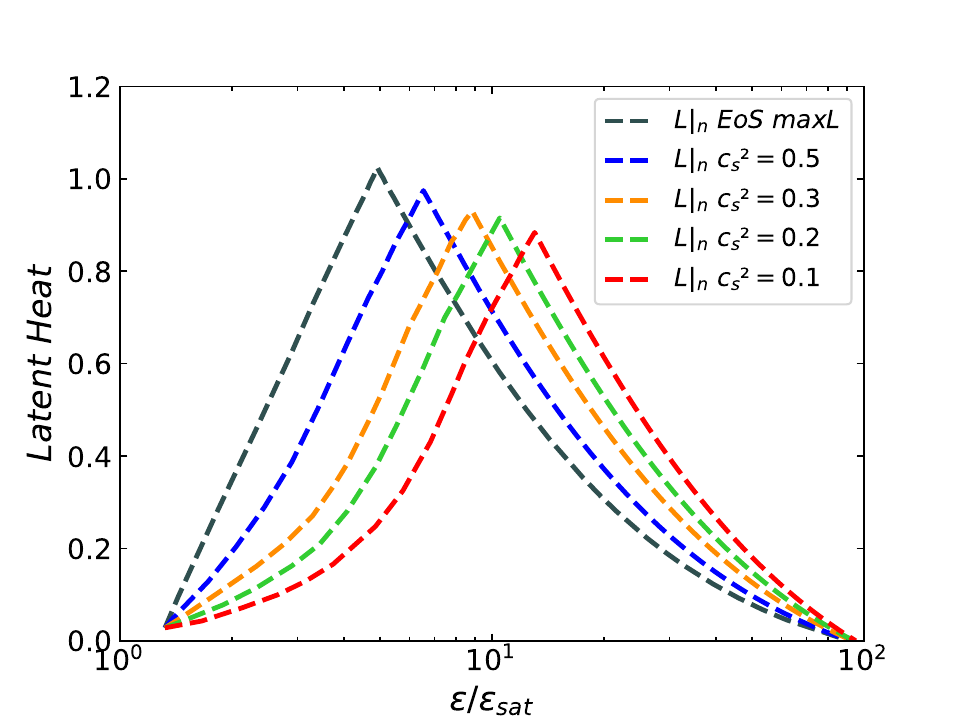}
\includegraphics[width=0.9\columnwidth]{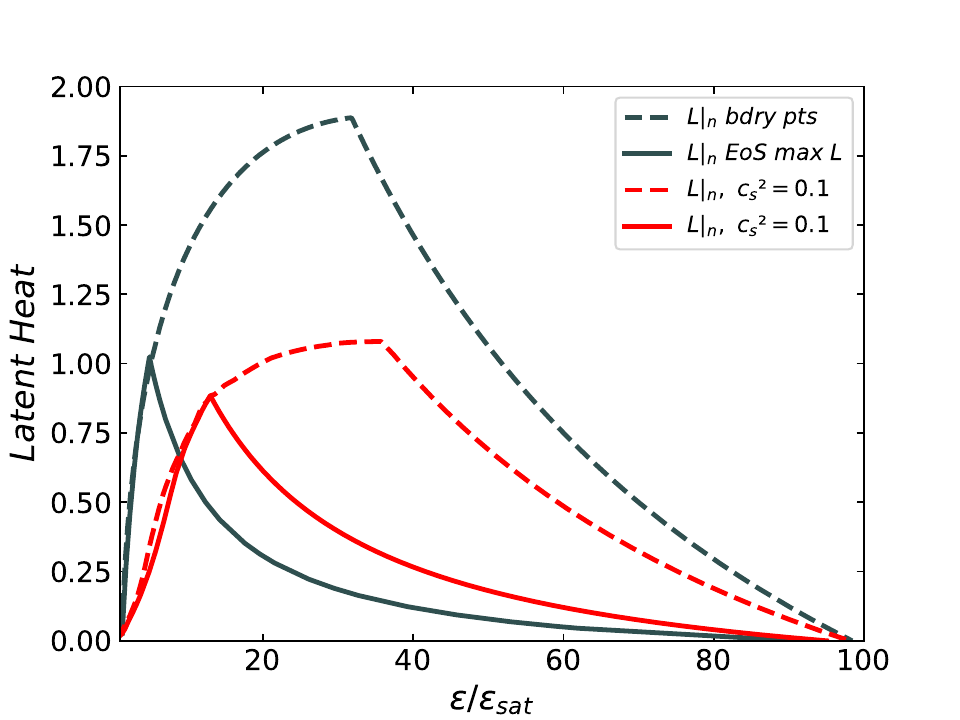}
\caption{{\bf Top:} Maximum specific latent heat $L|_n$ that fits within the nEoS band, with the same equations of state from figure~{fig:maxDE}  (top panel), but using Eq.~(\ref{diff}). The $OX$ axis (neutron number density at which the phase transition triggers) extends to 100$n_{\rm sat}$  (at chemical potential $\mu_B=2.6$ GeV) where  pQCD is matched.An absolute maximum $L|_n \simeq 1$ is reached for
$\varepsilon \simeq 5 \varepsilon_{\rm sat}$. 
{\bf Bottom:} Effect of including (solid lines) or not including (dashed lines) the integral constraints from the $(n,\mu)$ plane~\cite{Komoltsev:2021jzg}.
 \label{fig:L(n)}}
\end{figure}

\begin{figure}[h]
\includegraphics[width=0.9\columnwidth]{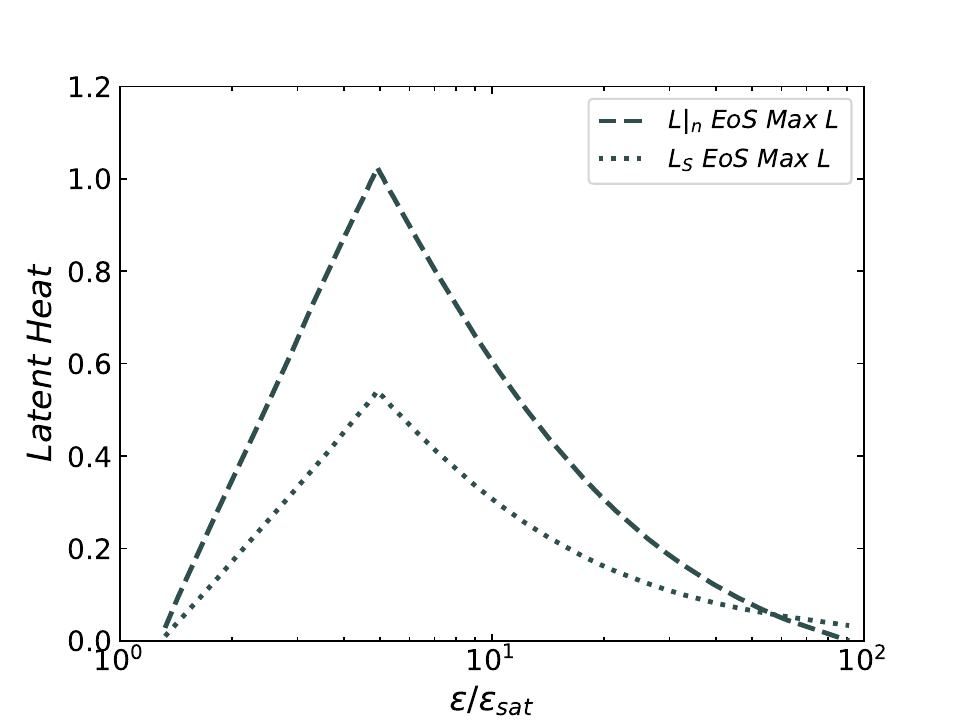}
\caption{We compare the bound on the latent heat $L|_n$ (upper line, dashed) from microscopic hadron physics alone (see fig.~\ref{fig:L}) with the maximum latent heat related to the Seidov limit~\cite{Seidov:1971sv}  (bottom line, dotted), by using Eq.~(\ref{diff}
\label{fig:compareSeidov2})}
\end{figure}

This is explicitly known for the ChPT low-density band due to complete nuclear calculations. For higher densities, where we are constructing all interpolating $P(\varepsilon)$, we need to obtain it iteratively from its lower-density values.
The two discretized equations to be stepped forward towards larger density (larger $i$ subindex) are
\begin{eqnarray}
n_i &=& \frac{\varepsilon_i}{M_N+(B/A)_i}\\
P_i &=& n^2_i \frac{(B/A)_{i+1}-(B/A)_{i-1}}{n_{i+1}-n_{i-1}}\ .
\end{eqnarray}
We use these two equations to solve for $(B/A)_{i+1}$
and $n_{i+1}$ given their values at the earlier two points $i$, $i-1$ and having our entire $P(\varepsilon)$ equation of state at hand. 
The system is started with the explicit low-density data from~\cite{Drischler:2016djf},  and the resulting $L|_n$ values are plotted in figures~\ref{fig:L(n)} and~\ref{fig:compareSeidov2}, in the same format as the earlier figures for $L|_\varepsilon$, namely \ref{fig:L} and \ref{fig:compareSeidov}.

Incidently, to transform the Seidov limit for $\Delta \varepsilon$ in Eq.~(\ref{Seidovsjump}) to a limit on the latent heat $L|_n$ as we plot it, we also need to express the number density $n = \varepsilon_{\rm max}^{\rm GR} /(M_N+B/A|_{\varepsilon_{\rm max}^{\rm GR}})$
in terms of the maximum energy density computed within General Relativity, equation
which is likewise iteratively stepped forward.

\newpage
\section{Discussion}

The specific latent heat, be it $L|_n$ of Eq.~(\ref{def:Ln}) or $L|_\varepsilon$ of Eq.~(\ref{LperA}), is a sensible quantity to characterize the intensity of a phase transition, but by no means the only one; for example, Lindblom~\cite{Lindblom:1998dp}
chooses to employ the also dimensionless
\begin{equation} \label{eqLindblom}
\Delta = \frac{\varepsilon_E-\varepsilon_H}{P_H+\varepsilon_H}\ .
\end{equation}
Yet a different choice is that of Seidov, who uses 
\begin{equation}
q-1:=\frac{\varepsilon_E-\varepsilon_H}{\varepsilon_H}
\end{equation}
normalizing the phase-transition discontinuity in $\varepsilon$ to the energy density at the low-end (hadronic)  $\varepsilon_H$. Since $P_H\ll \varepsilon_H$ in the explored regime of low-density physics, $\Delta$ and $q-1$ are almost proportional and not that different.

These or similar quantities share with our Eq.~(\ref{LperA}) the advantage of not being normalized to the hadron scale (unlike the nonrelativistic limit Eq.~(\ref{def:Ln}) with 
$L\propto M_N^{-1}$) but being usable for any problem. 
The definition that we have chosen, Eq.~(\ref{LperA}) is close to Eq.~(\ref{def:Ln})
for much of the span of neutron star physics in the MeV-GeV range, which immediately connects it  with many other subbranches of physics, whereas Lindblom's $\Delta$ in Eq.~(\ref{eqLindblom}) is less widely used.

The reader will have noticed in table~\ref{tab:L} that the entry corresponding to neutron star matter stands out in size: indeed, if a first order phase transition would be experimentally discovered in nuclear matter under pressure, it would hold the record across all known substances. This may happen indirectly in neutron stars, or as seems more likely, directly in relativistic heavy ion collisions below the presumed critical point~\cite{Shuryak:2020yrs}. 
(There are reasons to suspect an even higher-$L$ phase transition above the electroweak scale~\cite{Cline:2000fh}, to provide a nonequilibrium environment for baryogenesis, but there is  at the present time no clear experimental path to its discovery.)

The main result of this work is the specific latent heat reported in figures~\ref{fig:L} and~\ref{fig:Lfinal}. 
It is noteworthy that these curves there  represent the maximum possible such $L$ tolerated by the theory of nuclear strong interactions, in its present state, should the phase transition to an exotic QCD phase trigger at the corresponding $\varepsilon$ along each EoS of figure~\ref{fig:EoS}. We have sampled many more equations of state (the nEoS sets provide several thousands) but found no additional information.

In recent work~\cite{Lope-Oter:2021vxl} we show the effect of typical achievable temperatures on the neutron star EoS uncertainty band. Where those uncertainties are largest, $T\ll \rho$ and temperature does not play an important role, except for very superficial layers (that do not contribute much to the latent-heat bound)

Improvements in QCD theory or its low-density effective theory would shrink the nEoS band (gray and red shaded areas in figure~\ref{fig:EoS}) and therefore the maximum possible latent heat. In conclusion, we propose the function $L|_\varepsilon(\epsilon_H)$  as a diagnostic to quickly quantify future progress in constraining the equation of state of neutron star matter from first principles.
\vspace{0.2cm}

\emph{{\bf Acknowledgment:} 
Supported by grants MICINN: PID2019-108655GB-I00, PID2019-106080GB-C21 (Spain);
the COST action CA16214 (Multimessenger Physics and Astrophysics of Neutron Stars);
Univ. Complutense de Madrid under research group 910309 and the IPARCOS
institute. 
}

\newpage 



\end{document}